\documentclass[aps,pra,twocolumn,nopacs,floatfix]{revtex4-2}
\usepackage{epsfig}
\usepackage{graphicx}
\usepackage{dcolumn}
\usepackage{amsthm,amsmath}
\usepackage{comment}
\usepackage[colorlinks]{hyperref}
\usepackage{xcolor}
\usepackage[T1]{fontenc}

\usepackage{mathrsfs}
\usepackage{longtable}
\usepackage{orcidlink}
\usepackage{amsmath}
\usepackage[normalem]{ulem}

\usepackage{resizegather}
\UseRawInputEncoding
\begin{document}

\title{Relativistic equation-of-motion coupled-cluster theory analysis of black-body radiation shift in the clock transition of Zn I}

\author{Somesh Chamoli}

\author{Anmol Mishra}

\affiliation{Department of Chemistry, Indian Institute of Technology Bombay, Powai, Mumbai 400076, India}

\author{Richa Sharma Kesarkar}
\affiliation{Space Applications Centre, Indian Space Research Organization, Department of Space, Government of India, Ahmedabad, Gujarat 380058, India}
\author{B. K. Sahoo}
\email{bijaya@prl.res.in}
\affiliation{Atomic, Molecular and Optical Physics Division, Physical Research Laboratory, Navrangpura, Ahmedabad 380009, India}
\author{Achintya Kumar Dutta}
\email{achintya@chem.iitb.ac.in}
\affiliation{Department of Chemistry, Indian Institute of Technology Bombay, Powai, Mumbai 400076, India}

\begin{abstract}
   We have employed equation-of-motion coupled-cluster (EOM-CC) method in the four-component relativistic theory framework to understand roles of electron correlation effects in the {\it ab initio} estimations of electric dipole polarizabilities ($\alpha$) of the states engaged in the clock transition ($^{1}$S$_{0}$$\rightarrow$$^{3}$P$_{0}$) of the zinc atom. Roles of basis size, inclusion of higher-level excitations, and higher-order relativistic effects in the evaluation of both excitation energies of a few low-lying excited states and $\alpha$ are analyzed systematically. Our EOM-CC values are compared with the earlier reported theoretical and experimental results. This demonstrates the capability of the EOM-CC method to ascertain the preciseness of the black-body radiation shift in a clock transition, which holds paramount importance for optical clock-based experiments. 
\end{abstract}

\maketitle
\section{    Introduction}

An in-depth understanding of atomic properties that dictate an atom$'$s reaction to an external perturbation has attracted the attention of generations of theoreticians and experimentalists. Quantum mechanical calculations for accurate estimations of these properties have witnessed a significant surge in recent times due to their crucial applications in high-precision experiments \cite{MANAKOV1986320,doi:10.1142/2962,pethick_smith_2008}, particularly in the realm of atomic clocks \cite{Sahoo2017}. Atomic clocks are considered the most accurate time-keeping instruments on Earth today \cite{9316270,doi:10.1126/science.1102330,10.1063/1.2812121}. They are at the heart of all satellite-based navigation systems, i.e., the GNSS (Global Navigation Satellite System) \cite{jaduszliwer2021past}, including deep space navigation \cite{burt2021demonstration,10115636}, and also find their utility in quantum computing \cite{10.1063/PT.3.3626,brown2016co-designing} and telecommunication networks \cite{Lewandowski_2011,komar2014quantum}. Based on their frequency of operation, atomic clocks can be divided into two broad categories: microwave and optical clocks. The most common and well-understood atomic clocks operate in the microwave regime are based on the Cs (cesium) and Rb (rubidium) atoms, and also the H (hydrogen) maser \cite{doi:10.1126/science.1102330}. Their stabilities are inherently limited owing to their frequencies, and thus, in order to achieve higher stabilities, exploring the optical transitions has become the better possible bet \cite{vanier2015quantum}. The present-day state-of-the-art optical clocks can reach stabilities of the order of (10\textsuperscript{-21}) \cite{PhysRevLett.123.033201,sanner2019optical, Bothwell_2019}, and this number is ever decreasing with proper understanding of the systematic effects and development of adequate equipment \cite{bothwell2022resolving}.
In comparison to the most precise microwave clocks, these optical clocks promise comparable or better natural line widths, allowing for orders of magnitudes better performance in terms of oscillator quality factors and, hence, clock accuracy \cite{RevModPhys.87.637}.

The most modern atomic clocks will lose only one second in 300 billion years \cite{Zheng2022,bothwell2022resolving}. Since the atomic clocks rely on the fundamental interactions between the elementary particles for their operation, their output frequency values must remain the same irrespective of their location or time of measurement. However, in practice, they exhibit instability and reproducibility limitations due to environmental perturbations, which can shift their measured frequencies from their unperturbed natural atomic frequencies \cite{nicholson2015systematic,mcgrew2018atomic}.  This, in effect, can result in uncertainties in their output frequencies and can mislead the subsequent applications; for instance, it can result in major issues during navigation \cite{1572824499956313472}. It is of utmost importance to estimate the systematic effects in the optical clock frequency measurements. This demands performing additional precise measurements to study the sensitivity of atoms with external perturbations, which can also be challenging. On the other hand, adequate many-body methods that are capable of incorporating both electron correlation effects and relativistic corrections can help estimate them at a much lower cost.  

Among various quantum mechanical methods known today, the relativistic coupled-cluster (RCC) theory \cite{Liu2021RelativisticMethods,Kaldor2002RELATIVISTICELEMENTS} is renowned for its balanced inclusion of both relativistic and electron correlation effects, making it a potent and versatile approach. In this theory, major relativistic effects can be accounted for using the Dirac-Coulomb (DC) Hamiltonian. The accuracy of these calculations can be further enhanced by including higher-order relativistic effects, such as frequency-independent Breit interaction and leading-order quantum electrodynamics (QED) corrections.

One of the major sources of clock jumps in ultra-precision optical atomic clocks is the black-body radiation (BBR) shift of the clock transition caused by the interaction of the atomic system with thermal photons that make up the thermodynamic environment. In most cases, the BBR shift at room temperature T $\sim$ 300 K has a relative magnitude of 10\textsuperscript{-15}-10\textsuperscript{-16} with respect to the frequency of the reference optical transition, and the variation in BBR shift is a major obstacle to reaching the 10\textsuperscript{-19} level of stability and beyond. Prior to this, extensive, precise theoretical calculations were conducted to determine the BBR shifts for several potential atomic clocks \cite{PhysRevA.90.042505,PhysRevA.106.032807,Sahoo2017,PhysRevLett.97.040801,PhysRevA.80.062506,PhysRevA.83.030503,PhysRevA.85.012506}. However, only limited studies have been carried out for the Zn atom \cite{Dzuba_2019,PhysRevA.93.043420}. Precise measurement of the BBR shift for the Zn atom's clock transition remains elusive and lacks reliable determination to date. Ovsiannikov et al. \cite {PhysRevA.93.043420} calculated the BBR shift for the Zn atom using a model potential method. On the other hand, Dzuba et al. \cite{Dzuba_2019} determined the fractional BBR shift by using the parametrized configuration-interaction method combined with the many-body perturbation theory (CI+MBPT2) \cite{PhysRevA.54.3948} scaled with parameter fitted to the experimental energies. The truncated CI method suffers from the problem of size extensivity \cite{shavitt_bartlett_2009}; i.e., the total energy does not scale linearly with the number of electrons. Moreover, the MBPT2 method cannot give a proper description for heavy atoms, where valence electrons strongly interact with core-electrons \cite{atoms9040104}. One can partly mitigate the problem using the parametric CI+MBPT2 approach, which uses parameters fitted to the experimental data. However, such an approach would require the existence of prior experimental data and may not be equally accurate for all the properties. The equation-of-motion coupled-cluster (EOM-CC) method, on the other hand, is size extensive even in its truncated form, systematically improvable, and an {\it ab initio} approach; i.e., does not require any input from experiments and can provide high-accuracy results even in the heavier systems. In this work, we present a computational protocol for estimating the BBR shift of atomic clock candidates using the equation-of-motion approach based RCC (EOM-CC) method and Zn atom as a test case. We also analyze the effects of the basis-set size, higher-level excitations, and higher-order relativistic corrections in the determination of the BBR shift.

\section{Theory}

The BBR shift for a transition between the states $i$ and $j$ can be estimated using the approximation  \cite{PhysRevA.74.020502}
\begin{equation}
\label{eq:1}
\delta E= -\dfrac{1}{2}\left( 831.9\: V/m \right)^{2}\left( \dfrac{T}{300}\right)^{4}\left( \alpha^{0}_{i}-\alpha^{0}_{j}\right)\left( 1+\eta\right) ,
\end{equation}
where $\alpha^{0}_{i}$ and $\alpha^{0}_{j}$ are the static scalar polarizabilities of the $i$ and $j$ states, respectively, and $\eta$ is a small dynamic correction resulting from the frequency distribution of the BBR field, which can be neglected for the present interest of accuracy in the result. The static polarizability can be calculated using quantum mechanical many-body methods. In the context of perturbation theory, the energy associated with a specific state of an atom, when subjected to the influence of an external weak electric field with strength $\textit{\textbf{F}}$, can be described as \cite{MANAKOV1986320,doi:10.1142/2962}
\begin{equation}
\label{eq:2}
E(|\textit{\textbf{F}}|)=E(0)-\frac{\alpha}{2}|\textit{\textbf{F}}|^{2}-\cdots .
\end{equation}
In the above expression, $E(0)$ represents the energy of the state without any external electric field, and  $\alpha$ corresponds to the dipole polarizability of that particular state. By examining Eq. (\ref{eq:2}), it becomes evident that $\alpha$ can be ascertained through the assessment of the second-order differentiation of $E(|\textit{\textbf{F}}|)$ with respect to a small electric field of magnitude, $\textit{\textbf{F}}$. i.e.
\begin{equation}
\label{eq:3}
\alpha=-\left(\frac{\delta^{2}E(|\textit{\textbf{F}}|)}{\delta|\textit{\textbf{F}}|\delta|\textit{\textbf{F}}|}\right)_{|\textit{\textbf{F}}|=0} .
\end{equation}
The energy derivative presented in Eq. (\ref{eq:3}) can be computed through both analytical and numerical approaches. The process of calculating the derivative using an analytical approach, also known as the analytical gradient method, necessitates the differentiation of the energy functional with respect to the external perturbation (electric field in the present case), which can be tedious for the RCC method and it is neither variational nor hermitian. The alternative technique, referred to as the finite-field approach, entails numerically evaluating the energy derivative at various electric field strengths. By applying the finite-field (FF) method, the energy eigenvalues are obtained by perturbing the atomic Hamiltonian with the interaction Hamiltonian ($H^{'} = -\textit{\textbf{F}}\cdot{\textit{\textbf{D}}}$), where $\textit{\textbf{D}}$ represents the induced electric dipole moment in the perturbative Hamiltonian. The FF approach offers a significant benefit as it effortlessly extends to analyze the characteristics of both the ground and excited states of a system, and the calculations at different field strengths can be performed in parallel. In the course of our study, we employed numerical differentiation to calculate the polarizabilities. We computed the total energies twice, once with perturbation and once without. Throughout the calculations, we maintained a constant electric field value of 0.001 atomic units (a.u.).

\section{Method for calculation}

\subsection{Relativistic EOM-CC method}

In the (R)CC theory formalism, the wavefunction of the ground state of an atomic system is expressed as
\begin{equation}
\label{eq:4}
|\Psi_{cc}\rangle=e^{\hat{T}}|\Phi_{0}\rangle ,
\end{equation}
where $|\Phi_{0} \rangle$ denotes the reference determinant, and $\hat{T}$ is the RCC operator that generates different levels of excitations from the reference state. These excitation levels are denoted by subscripts as
\begin{equation}
\label{eq:5}
\hat{T}=\hat{T_{1}}+\hat{T_{2}}+ \cdots + \hat{T_{N}} ,
\end{equation}
where any general n-tuple excitation RCC operator can be expressed as
\begin{equation}
\label{eq:6}
\hat{T_{n}}=\left(\frac{1}{n!}\right)^{2}\sum_{ij...ab...}^{n}{t_{ij...}^{ab...}}a_{a}^{\dagger}a_{b}^{\dagger}...\hspace{0.2cm}a_{j}a_{i} \cdots .
\end{equation}
In Eq. (\ref{eq:6}), $t_{ij...}^{ab...}$ are the cluster amplitudes, indices $\left(i\hspace{0.1cm},j \hspace{0.1cm},k\hspace{0.1cm}...\right)$ represent occupied spinors and virtual spinors are shown by $\left(a\hspace{0.1cm},b\hspace{0.1cm},c\hspace{0.1cm}...\right)$. The reference state $|\Phi_{0}\rangle$ is obtained using the Dirac-Hartree-Fock (DHF) method. The commonly used singles-doubles approximation in the RCC (CCSD) method is achieved by limiting the cluster operator to include solely one-body and two-body excitations. The RCC amplitudes are obtained by solving a set of simultaneous non-linear equations
\begin{equation}
\label{eq:7}
\langle \Phi_{ij...}^{ab...}|\Bar{H}|\Phi_{0}\rangle=0 ,
\end{equation}
where $ | \Phi_{ij..}^{ab..} \rangle$ are  the excited determinants and
$\Bar{H}=e^{-\hat{T}}\hat{H}e^{\hat{T}}$ is the similarity transformed Hamiltonian. $\hat{H}$ represents the DC Hamiltonian, which is defined as
\begin{equation}
\label{eq:8}
\hat{H}=\sum_i^N\left[c \boldsymbol{\alpha_i} \cdot \boldsymbol{p_i}+\beta_i m_0 c^2+V_{nuc}(r_i)\right]+\sum_{i>j}^N \frac{1}{r_{i j}} .
\end{equation}
In Eq. (\ref{eq:8}), $\boldsymbol{\alpha}$ and $\beta$ are the Dirac matrices, $V_{nuc}(r)$ represents the nuclear potential, $m_{0}$ is the rest mass of a free electron and $c$ stands for the speed of light. Eq. (\ref{eq:8}) considers only positive energy electrons within the summation while also ensuring the accurate implementation of the no-pair approximation. The DC Hamiltonian can be augmented by incorporating the Gaunt term, which considers electromagnetic interactions, or the Breit term, which accounts for both the electromagnetic interactions and electron retardation effects,
\begin{equation}
\label{eq:9}
\hat{H}=\sum_i^N\left[c \boldsymbol{\alpha_i} \cdot \boldsymbol{p_i}+\beta_i m_0 c^2+V_{nuc}(r_i)\right]+\sum_{i>j}^N\left(\frac{1}{r_{i j}}+B_{i j}\right) I_4 ,
\end{equation}
where 
\begin{equation}
\label{eq:10}
B_{i j}=-\frac{1}{2 r_{i j}}\left[\boldsymbol{\alpha_i} \cdot \boldsymbol{\alpha_j}+\frac{\left(\boldsymbol{\alpha_i} \cdot \boldsymbol{r_{i j}}\right)\left(\boldsymbol{\alpha_j} \cdot \boldsymbol{r_{i j}}\right)}{r_{i j}^2}\right]
\end{equation}
is the Breit operator with, 
\begin{equation}
\label{eq:11}
G_{i j}=-\frac{\boldsymbol{\alpha_i} \cdot \boldsymbol{\alpha_j}}{2 r_{i j}}
\end{equation}
as a Gaunt term and 
\begin{equation}
\label{eq:12}
-\frac{1}{2 r_{i j}}\left[\frac{\left(\boldsymbol{\alpha_i} \cdot \boldsymbol{r_{i j}}\right)\left(\boldsymbol{\alpha_j} \cdot \boldsymbol{r_{i j}}\right)}{r_{i j}^2}\right] 
\end{equation}
as retardation effect.

The ground state energy expression in the RCC theory is given as
\begin{equation}
\label{eq:13}
\langle \Phi_{0}|\Bar{H}|\Phi_{0}\rangle=E .
\end{equation}
The CCSD method approximation frequently falls short in delivering precise quantitative accuracy, prompting us to incorporate additional higher-level excitations in the form of triples (CCSDT), quadruples (CCSDTQ), and pentuples (CCSDTQP) to achieve more accurate results of interest. However, incorporating higher excitation comes with a  significantly higher computational cost. 

The RCC method is extended to obtain the excited states using the EOM-CC approach \cite{ROWE1968,MUKHERJEE1979325,10.1063/1.464746}. In the EOM-CC method, the $k^{th}$ excited state can be expressed as
\begin{equation}
\label{eq:15}
|\Psi_{k}\rangle=\hat{R}_{k}e^{\hat{T}}|\Phi_{0}\rangle ,
\end{equation}
where $\hat{R}_{k}$ is a linear excitation operator with a form of
\begin{equation}
\label{eq:16}
\hat{R}_{k}= r_{0}+\sum_{i,a} r_{i}^{a}\;\hat{a}_{a}^{\dagger}\hat{a}_{i}+\sum_{i<j\atop a < b} r_{i j}^{ab}\;\hat{a}_a^{\dagger}\hat{a}_b^{\dagger} \hat{a}_j \hat{a}_{i}+\cdots .
\end{equation}
In Eq. (\ref{eq:16}), $r_{i}^{a}$, $r_{ij}^{ab}$ $\cdots$ are the amplitudes corresponding to singly, doubly, and higher excited configurations and $r_{0}$ is a constant denoting the overlap with the ground state. The Dirac equation for the $k^{th}$ excited state in the EOM-CC picture can be written as 
\begin{equation}
\label{eq:17}
\hat{H}\hat{R}_{k}e^{\hat{T}}|\Phi_{0}\rangle=E_{k}\hat{R}_{k}e^{\hat{T}}|\Phi_{0}\rangle .
\end{equation}
After following some mathematical algebra and using the commutative property of the $\hat{R}_{k}$ and $\hat{T}$ operators, we arrive at
\begin{equation}
\label{eq:18}
\bar{H}\hat{R}_{k}|\Phi_{0}\rangle=E_{k}\hat{R}_{k}|\Phi_{0}\rangle
\end{equation}
and
\begin{equation}
\label{eq:19}
\hat{R}_{k}\bar{H}|\Phi_{0}\rangle=E_{0}\hat{R}_{k}|\Phi_{0}\rangle .
\end{equation}
Taking difference of Eq. (\ref{eq:18}) and Eq. (\ref{eq:19}), we get
\begin{equation}
\label{eq:20}
[\bar{H} , \hat{R}_{k}]|\Phi_{0}\rangle=\omega_{k}\hat{R}_{k}|\Phi_{0}\rangle .
\end{equation}
Here $\omega_{k}=E_{k}-E_{0}$ is our excitation energy. Similar to the CCSD method, the EOM-CC equations are generally truncated at the singles and doubles approximation, resulting in the EOM-CCSD method. Davidson's iterative diagonalization method \cite{Hirao1982} is typically employed to solve the EOM-CC equation.
Because of the non-Hermitian nature of $\Bar{H}$, it also exhibits a left eigenvector with the eigenvalue equation as
\begin{equation}
\label{eq:21}
\langle \Phi_{0}|\hat L_{k}\bar{H} = \langle \Phi_{0}|\hat L_{k} E_{k} ,
\end{equation}
where $\hat{L}_{k}$ is a linear de-excitation operator with a form of
\begin{equation}
\label{eq:22}
\hat{L}_{k}= l_{0}+\sum_{i,a} l_{a}^{i}\;\hat{a}_{i}^{\dagger}\hat{a}_{a}+\sum_{i<j\atop a < b} l_{ab}^{ij}\;\hat{a}_i^{\dagger}\hat{a}_j^{\dagger} \hat{a}_b \hat{a}_{a}+\cdots .
\end{equation}
The two sets of eigenfunctions follow the biorthogonality condition, such as
\begin{equation}
\label{eq:23}
\langle \Phi_{0}|\hat L_{k}\hat{R}_{l}|\Phi_{0} \rangle = \delta_{kl} .
\end{equation}
For the calculation of energy, it is sufficient to solve either of the left or right eigenvalue equations, whereas an analytical calculation of properties requires a solution of both left and right eigenvectors. As mentioned before, the EOM-CCSD method frequently falls short in delivering precise quantitative accuracy. So, it demands to incorporate additional higher-level excited configurations such as triples (EOM-CCSDT), quadruples (EOM-CCSDTQ), and pentuples (EOM-CCSDTQP) both in the ground and excited states to achieve more accurate values for the calculated energies and other properties.
The computational cost in the EOM-CCSDT, EOM-CCSDTQ, and EOM-CCSDTQP methods scale as $O(N^{8})$, $O(N^{10})$ and  $O(N^{12})$, where $N$ is the typical number of basis functions used in the calculation. All the relativistic EOM-CC calculations are done using the DIRAC \cite{DIRAC23} interface of the MRCC software \cite{10.1063/1.5142048}. The relativistic calculations are performed using the four-component DC Hamiltonian unless explicitly mentioned otherwise. 

\begin{table}[t]
\centering
\caption{Excitation energies of the zinc atom calculated using the at aug-cc-pVTZ basis set in the relativistic EOM-CCSD method. All the core electrons are used to account for the correlation effects. The results are compared with the available experimental and previously reported theoretical values.}
\begin{tabular}{ c c c c}
\hline \hline \\
  & \multicolumn{3}{c}{Excitation Energy  (in cm\textsuperscript{-1 })}   \\
 \cline{2-4}  \\ State & This work    & Experiment \cite{DagGullberg_2000} & Previous work \cite{Dzuba_2019}  \\
 \hline \\
 4s4p $^{3}$P$_{0}$  & 31452     & 32311   & 32348   \\
 4s4p $^{3}$P$_{1}$  & 31641     & 32501   & 32546   \\
 4s4p $^{3}$P$_{2}$  & 32025     & 32890   & 32950    \\
 4s4p $^{1}$P$_{1}$  & 46703     & 46745  & 46908     \\
\hline \hline 
\end{tabular}
\label{table:1}
\end{table}

\section{Results and discussion}

Table \ref{table:1} displays the excitation energy values for the Zn atom's first four excited states; 4s4p ($^{3}$P$_{0}$, $^{3}$P$_{1}$, $^{3}$P$_{2}$, and $^{1}$P$_{1}$). These values were computed using the relativistic EOM-CCSD method with an uncontracted aug-cc-pVTZ basis set. The calculation was performed considering all-electron correlation effects. Table \ref{table:1} also includes a compilation of excitation energies for the Zn atom that has been reported in earlier theoretical study \cite{Dzuba_2019} and experiment \cite{DagGullberg_2000}. The scaled approach used in Ref. \cite{Dzuba_2019} overestimates the excitation energies with respect to the experimental values. For the triplet states, there is a discrepancy of 37${-}$60 cm\textsuperscript{-1} compared to the experimental data, while the first singlet state ($^{1}$P$_{1}$) exhibits a larger deviation of 163 cm\textsuperscript{-1}.
Despite scaling with the experimental data, the estimated value still exhibits disagreement with the observed results. In comparison to the experimental values, our results are significantly underestimated. 
The discrepancies in the excitation energy values of the first three triplet states are approximately within the range of 859 to 865 cm\textsuperscript{-1}, whereas for the $^{1}$P$_{1}$ singlet state, the difference is much smaller, at only about 42 cm\textsuperscript{-1}. The reasons for such discrepancies in an {\it ab initio} approach are discussed below.

\begin{figure}[t]
    \begin{center}
    \includegraphics[width=0.5\textwidth]{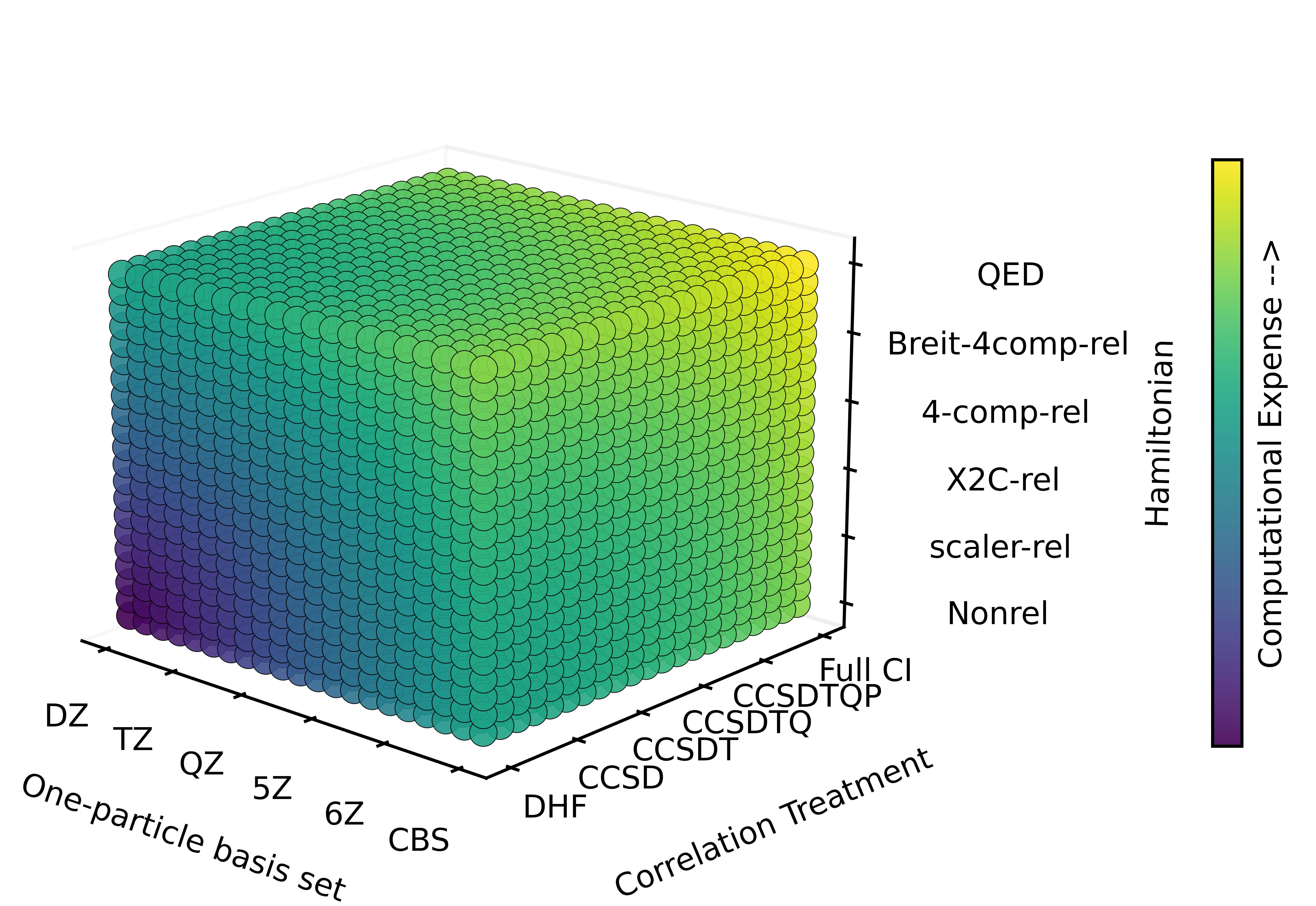}
    \caption{Demonstration of the factors deciding accuracy of a property in the multi-electronic atom using the quantum chemical approach.}
    \label{fig:my_label1}
    \end{center}
\end{figure}

The accuracy of calculated energy, for that matter any property, in the {\it ab initio} approach of a many-body method in an atomic system relies on three pillars. Accuracy of the correlation method, completeness of one electron basis set used, and the nature of the Hamiltonian used. In an ideal case, one needs to get convergence with respect to all three factors (See Fig. \ref{fig:my_label1}). It is intended to have a relativistic Hamiltonian, which includes full quantum electrodynamics, to get the best possible accuracy. However, in practice, using the four-component DC Hamiltonian is often sufficient with the correction included for Gaunt or Breit term for atomic clock application. We have used the EOM-CC method to account for the electron correlation effects more rigorously. One can improve the accuracy of the calculated results, including triply or even higher-level excited state configurations in the calculations. It is one of the advantages of the RCC method that the results can be systematically improved by including a higher-level excited cluster operator, of course at the expense of higher computational cost. However, the use of very accurate wavefunction-based methods like RCC theory leads to very slow convergence of the results with respect to the size of one electron basis set. This work considers all these three effects in a sequential manner, restricting our attention only to the clock transition state.

\subsection{Effect of basis functions}

Table \ref{table:2} presents the polarizability values for the ground state ($^{1}$S$_{0}$), the first excited triplet state ($^{3}$P$_{0}$), and the excitation energies related to the clock transition ($^{1}$S$_{0}$$\rightarrow$$^{3}$P$_{0}$) of the Zn atom at different basis sets. We have performed the calculations using an uncontracted aug-cc-pVXZ (X=D, T, Q, and P) basis set with a frozen core approximation (keeping ten core electrons frozen for the calculation) to investigate if the deviation in results could be attributed to the utilization of a finite-size basis set. 

The analysis of Table \ref{table:2} reveals a clear pattern for the excitation energy: as we move up the hierarchy of the basis set, the excitation energy decreases. The change from aug-cc-pVDZ to aug-cc-pVTZ is 572 cm\textsuperscript{-1 }. The change is smaller at 42 cm\textsuperscript{-1 } on going from aug-cc-pVTZ to aug-cc-pVQZ. The excitation energy does not converge even at the aug-cc-pVQZ basis set, and moving to the aug-cc-pV5Z basis set leads to a change of 24 cm\textsuperscript{-1 }. The extrapolation to the complete basis set (CBS) limit using the Martin extrapolation formula \cite{MARTIN1996669} leads to an excitation energy of 31578 cm\textsuperscript{-1 }. The CBS extrapolation can be a major source of potential error and will be discussed in detail in the error estimation section later. In contrast to the excitation energy, the polarizability of both the ground and excited states exhibit an erratic pattern as the number of basis functions increases. As we make the transition from the aug-cc-pVDZ to the aug-cc-pVTZ basis set, the polarizability value of the ground state decreases. However, upon further advancement to the aug-cc-pVQZ basis, a slight rise (0.11 a.u.) in the polarizability of the zinc atom's ground state is observed compared to the value obtained using the aug-cc-pVTZ basis. Increasing the number of basis functions in a basis set will decrease the absolute energy value due to the variational theorem. However, the validity of this argument cannot be guaranteed for properties like polarizability. The increase in the basis set to aug-cc-pV5Z basis leads to a further increase of 0.12 a.u.. The CBS level values for polarizability are obtained by extrapolating the total energy in the presence and absence of the electric field, and the CBS polarizability for the ground state is 39.73 a.u. In the case of the first excited state, we observe a similar trend, yet the decline from the double-zeta (DZ) to the triple-zeta (TZ) and the subsequent rise from TZ to quadruple-zeta (QZ) occurs at a much swifter rate compared to the ground state. The change from aug-cc-pVTZ to aug-cc-pVQZ is large at 6.84 a.u.. The subsequent change on going to aug-cc-pV5Z is smaller at 2.6 a.u.. However, the excited state polarizability did not converge even with the aug-cc-pV5Z basis set, and CBS extrapolation leads to a value of 64.82 a.u.. 

Therefore, the CBS extrapolation at the EOM-CCSD level approximation in the RCC theory gives discrepancies of 733 cm\textsuperscript{-1 } for the excitation energy and 0.93 a.u. for the ground state polarizability when compared with their respective experimental values. The remaining contribution is presumably due to the correlation effects arising from the higher-level excitations and omitted relativistic effects. The experimental estimate of the polarizability for the excited state is not available to make a comparative analysis. 
 
\begin{table}[t]
\centering
\caption{Comparison of the electric dipole polarizabilities of the states involved in the clock transition and its excitation energy in the Zn atom calculated using the relativistic EOM-CCSD method. The aug-cc-pVXZ (X=D, T, Q, and 5) basis sets are used for both the frozen-core (10 electrons) and complete basis set (CBS) limit approximations.}
\begin{tabular}{ c c c c}
\hline \hline \\
  & \multicolumn{2}{c}{Polarizability (in a.u.)}   \\
 \cline{2-3}  \\ Basis &$^{1}$S$_{0}$     & $^{3}$P$_{0}$ & Excitation energy \\
 &&&(in cm\textsuperscript{-1 })  \\
 \hline \\
 aug-cc-pVDZ    & 40.23    & 54.24    & 30906  \\
 aug-cc-pVTZ    & 39.30    & 52.75    & 31478  \\
 aug-cc-pVQZ    & 39.41    & 59.59    & 31520  \\
 aug-cc-pV5Z    & 39.53    & 62.19    & 31544  \\
 CBS-TZ/QZ/5Z   & 39.73    & 64.82    & 31578  \\
  \\
 Experiment  & 38.8(8)\cite{PhysRevA.54.1973}  & --   & 32311\cite{DagGullberg_2000} \\
\hline \hline 
\end{tabular}
\label{table:2}
\end{table}

\subsection{Roles of electron correlation effects}

To observe the impact of the absent correlation effects in our findings, we integrated the triples, quadruples, and pentuples excitations by conducting calculations using the CCSDT, CCSDTQ, and CCSDTQP methods. The calculation of the unaccounted correlation contributions to the $\alpha$ values and excitation energies is presented in the following way
\begin{equation}
\label{eq:24}
\Delta{T}_{(\alpha/EE)} = \text{CCSDT}_{(\alpha/EE)} - \text{CCSD}_{(\alpha/EE)} ,
\end{equation}
where the subscripts $\alpha/EE$ stand for the polarizability or excitation energy at a particular basis set arising from the triple excitations using the CCSDT method. Similarly, $\Delta{Q}$ and $\Delta{P}$ notations are used to mention the extra correlation contributions from the CCSDTQ and CCSDTQP methods, respectively. 

The uncontracted aug-cc-pVTZ basis set was employed for evaluating the $\Delta{T}$ correction with frozen core approximation. The $\Delta{Q}$ and $\Delta{P}$ corrections were assessed using the MRCC program in its non-relativistic version, employing the cc-pVDZ basis set with ten inner electrons kept frozen. Table \ref{table:3} clearly demonstrates that the inclusion of full triples has a more significant impact on both excitation energy and polarizabilities of both the ground and excited states. The inclusion of triples correction leads to an increase of 729 cm\textsuperscript{-1 } in the excitation energy. The ground state and the excited state polarizability decrease by 1.49 a.u. and 0.28 a.u., respectively.  Nevertheless, it is worth noting that the impact of quadruple excitations is significantly smaller compared to that of triples in the case of both polarizabilities and excitation energy. The excitation energy increases by 75 cm\textsuperscript{-1 } on the inclusion of quadruple excitation in the calculations. The ground and excited state polarizabilities increase by 0.04 and 0.23 a.u., respectively. The inclusion of pentuple excitation leads to a negligible change in the energy and polarizability values. The results can be presumed to be converged with respect to the electron correlation effects. The effects of sextuple and higher-level excitations can be safely disregarded in the present study.

\begin{table}[t]
\centering
\caption{Demonstration of the effects of the correlations and higher-order relativistic corrections to the CBS results of the excitation energy and polarizabilities of the Zn atom. Roles of the inner core electron correlations are also shown explicitly.}
\resizebox{0.5\textwidth}{!}{
\begin{tabular}{ c c c c}
\hline \hline \\
  & \multicolumn{2}{c}{Polarizability (in a.u.)}   \\
 \cline{2-3}  \\ Method &$^{1}$S$_{0}$     & $^{3}$P$_{0}$ & Excitation energy \\
 &&&(in cm\textsuperscript{-1 })  \\
 \hline \\
 CCSD   & 39.73  & 64.82  & 31578  \\
 (CBS--TZ/QZ/5Z) &&&\\
 \\
 + $\Delta{T}$ correction & - 1.49  & - 0.28   & + 729 \\
  (aug-cc-pVTZ)    &        &                 &   \\
  \\
 + $\Delta{Q}$ correction  & + 0.04    & + 0.23   & + 75 \\
  (non-rel)(cc-pVDZ) & & & \\
  \\
 + $\Delta{P}$ correction & + 0.03   & - 0.01     & - 10 \\
  (non-rel)(cc-pVDZ) & & & \\
  \\
 + $\Delta$core correction & + 0.01   & - 0.07     & - 20 \\
  (d-aug-dyall.ae2z/v2z) & & & \\
  \\
 + $\Delta$Gaunt correction & + 0.06               & + 0.12                & - 9 \\
  (aug-cc-pVDZ)    &        &                 &   \\
  \\
 Composite Value  & 38.38     & 64.81     &  32343 \\ \\

\hline \hline 
\end{tabular}
}
\label{table:3}
\end{table}

\subsection{Inner-core correlation and higher-order relativistic effects}

To reduce the computational cost of the calculation, we utilized the frozen core approximation throughout by compromising with accuracy of the results slightly. The aug-cc-pVXZ family of basis sets cannot take care of the core-correlation effects and are designed to be used with frozen core approximation \cite{10.1063/1.1520138}. Consequently, to assess the specific impact of correlation effects attributed to the frozen inner 10 electrons in our analysis, we carried out two sets of calculations. Firstly, we considered all 30 core electrons using an uncontracted d-aug-dyall.ae2z basis set, designed to take care of the core correlation. Secondly, we repeated the calculation with d-aug-dyall.v2z basis while freezing the inner 10 electrons. The difference in the results was considered as $\Delta$core and was added to our CBS results. 
It is evident from Table \ref{table:3} that the impact of the core electron correlations on the polarizability and excitation energy can be regarded as insignificant and are less than 0.3 percent and 0.1 percent for $\alpha$ and energy, respectively. It is also important to investigate the influence of higher-order relativistic corrections on the excitation energy. To account for the Gaunt contribution, the Hamiltonian is augmented by incorporating the Gaunt operator (as defined in Eq. (\ref{eq:11})). It should be noted that the Gaunt correction is added only at the level of DHF wavefunction evaluation, but it is not included in the inclusion of correlation effects. Subsequently, a relativistic EOM-CCSD calculation is executed using an uncontracted aug-cc-pVDZ basis. Again, repeating the same procedure but without adding the Gaunt operator to the Hamiltonian. By subtracting the properties of both calculations, we arrive at the $\Delta$Gaunt correction. Based on the data presented in Table \ref{table:3}, it is evident that the impact of Gaunt correction is not significantly pronounced. The Gaunt correction induces a redshift in the excitation energy, but its magnitude is so minimal that it can be regarded as negligible. This further affirms the insignificance of other higher-order relativistic corrections in the calculations for the Zn atom. Upon incorporating all the necessary corrections into the CBS result, we obtained a combined value of 32343 cm\textsuperscript{-1 } for excitation energy. The composite values for the ground and excited state polarizability are 38.38 a.u. and 64.81 a.u., respectively. The ground state polarizability shows a small 1.1$ \%$ deviation from the experiment and within the experimental error bar. The excitation energy shows a negligible error of 0.1\%. This implies high precision estimations of results in our {\it ab initio} approach.

\subsection{Error Estimation }

\begin{figure}[ht]
    \begin{center}
    \includegraphics[width=0.5\textwidth]{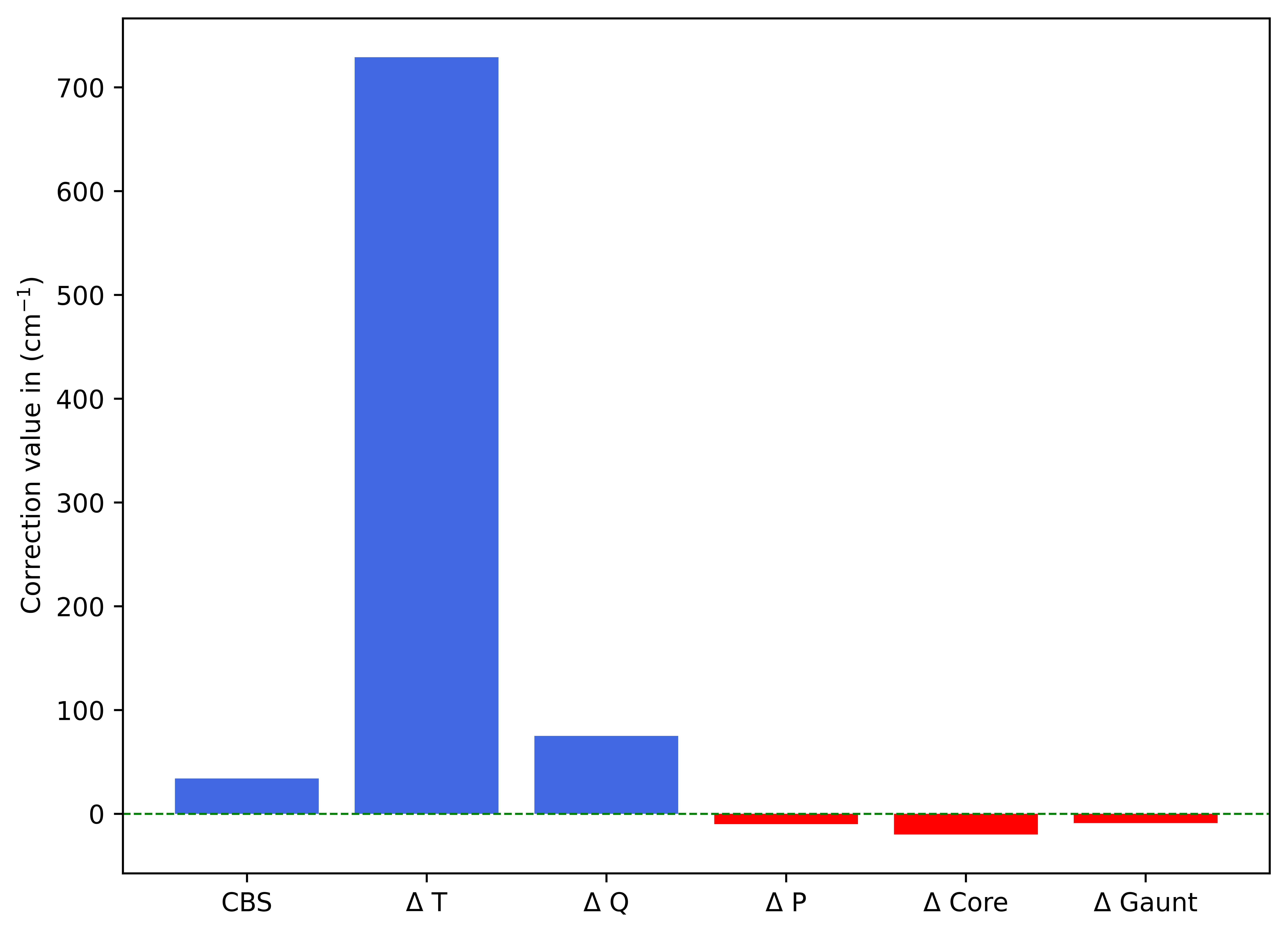}
    \caption{Effects of different corrections in the aug-cc-pV5Z results for the excitation energy of the Zn atom.}
    \label{fig:my_label2}
    \end{center}
\end{figure}

\begin{figure}[ht]
    \begin{center}
    \includegraphics[width=0.5\textwidth]{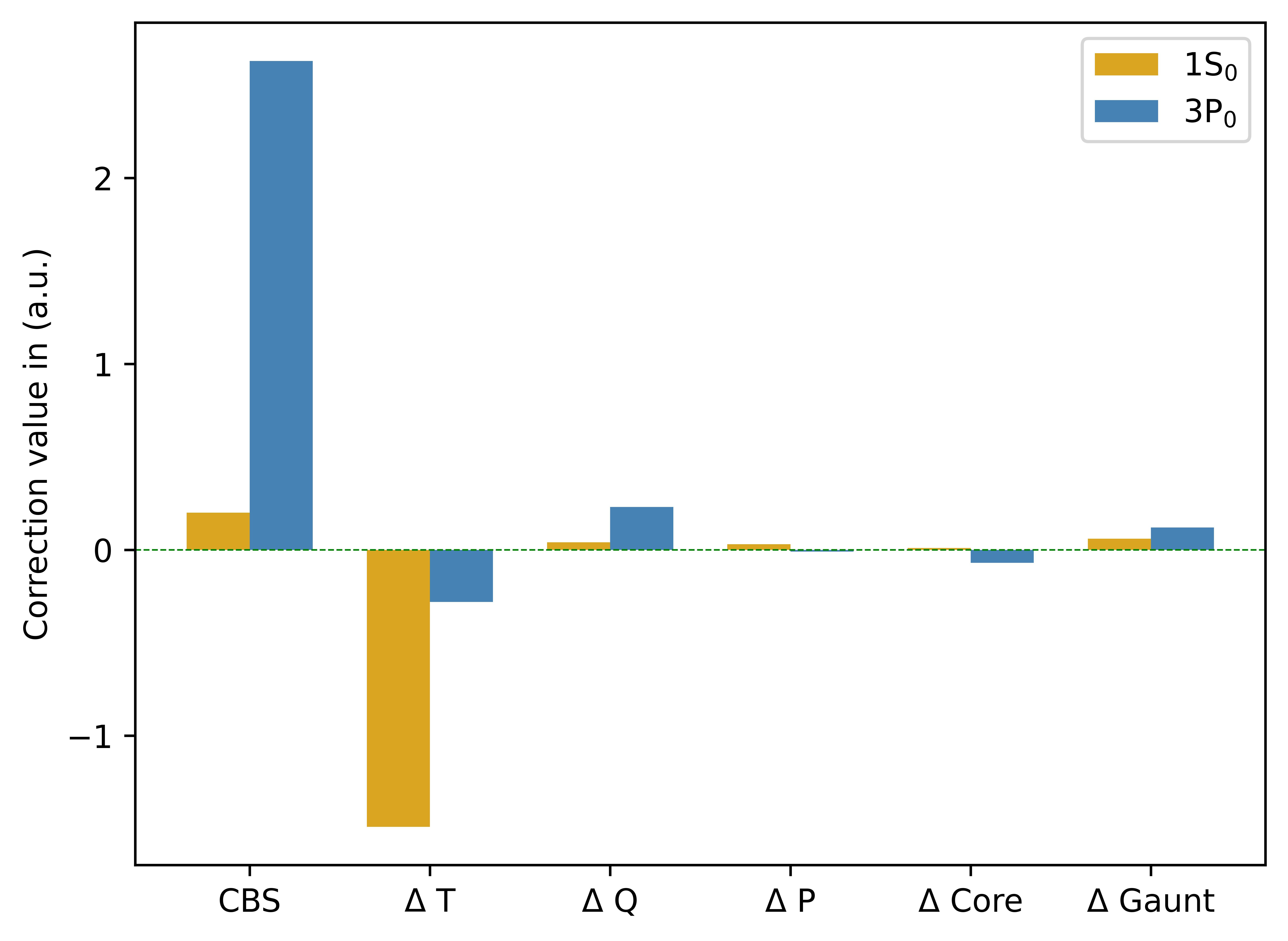}
    \caption{Effects of different corrections in the aug-cc-pV5Z results for the polarizability values of the ground ($^{1}$S$_{0}$) and first excited states ($^{3}$P$_{0}$) of the Zn atom.}
    \label{fig:my_label3}
    \end{center}
\end{figure}

Fig. \ref{fig:my_label2} represents the contribution of all different corrections added to the excitation energies from the uncontracted aug-cc-pV5Z basis.  It is clear that the major contribution comes from the triples, making its role pivotal for high-precision calculations. The effect of quadruple is small, and the pentuple excitation leads to a negligible value.  Fig. \ref{fig:my_label3} shows contributions from all possible corrections to the polarizability values of the ground and first excited states. It is clear from Fig. \ref{fig:my_label3} that obtaining a precise polarizability value for the excited state necessitates going beyond a finite-size basis. Also, the contribution from the triples on polarizability can not be ignored for both the ground and excited states. The non-additivity of the triples and higher-order correlation correction over different basis sets can be a source of uncertainty. The EOM-CCSDT values are available in two different basis sets -- cc-pVDZ and aug-cc-pVTZ basis sets. The difference of the excitation energy from the two basis sets in the CCSDT method is 736 cm\textsuperscript{-1}, which is of the same order as the difference observed in the EOM-CCSD level (744 cm\textsuperscript{-1}). The difference between the two has been considered as the uncertainty in $\Delta$ CCSDT excitation energy. Similarly, the uncertainty in the ground state and excited state polarizability for the $\Delta$ EOM-CCSDT method is 0.15 a.u. and 0.35 a.u., respectively. The results for the EOM-CCSDTQ and EOM-CCSDTQP methods are only available in a single basis set using a non-relativistic Hamiltonian. Earlier studies have shown that the error due to basis set incompleteness in the higher-level RCC method cannot be more than 50 percent of the value obtained using the smaller basis set \cite{10.1063/1.1811608}. Therefore, a conservative estimate will be half of the total correction at the $\Delta$ EOM-CCSDTQ and $\Delta$ EOM-CCSDTQP level for both excitation energy and polarizability.

\begin{table}[t]
\centering
\caption{Estimation of the uncertainty caused by various components.}
\resizebox{0.5\textwidth}{!}{
\begin{tabular}{ c c c c}
\hline \hline \\
  & \multicolumn{2}{c}{Polarizability (in a.u.)}   \\
 \cline{2-3}  \\ Contribution &$^{1}$S$_{0}$     & $^{3}$P$_{0}$ & Excitation energy \\
 &&&(in cm\textsuperscript{-1 })  \\
 \hline \\
$\Delta$ EOM-CCSDT & 0.15 & 0.35 & 8 \\
$\Delta$ EOM-CCSDTQ & 0.02 & 0.12 &38 \\
$\Delta$ EOM-CCSDTQP & 0.02 & 0.01 &5 \\
Missing higher order correlation & 0.03 & 0.01 & 10 \\
Basis set error in EOM-CCSD & 0.11 & 0.67 & 16\\
Core correlation & 0.01 & 0.04 & 10 \\
 Higher relativistic effect & 0.03 & 0.06 & 5\\
Final value   &  0.37   & 1.26     &  92 \\\\
\hline \hline 
\end{tabular}
}
\label{table:4}
\end{table}
The change caused by the pentuple excitation is negligible for both the estimations of the excitation energy and $\alpha$ values of the ground and excited states. One can consider the results to be converged with respect to the electron correlation effect on the inclusion of pentuple excited configuration in EOM-CC calculations. Therefore, the change caused by the inclusion of the pentuple excited configuration can be considered uncertainty due to the missing higher-order correlation effects. This would lead to 0.03 a.u. and 0.01 a.u., uncertainty for the ground and excited state polarizabilities, respectively. The uncertainty in the excitation energy will be ten cm\textsuperscript{-1 } due to missing higher-order correlations (see Table \ref{table:4}).

\begin{table}[t]
\centering
\caption{Effects of various basis set extrapolation schemes used to extrapolate the excitation energy and dipole polarizabilities. Deviations from the aug-cc-pV5Z values are provided in the bracket.}
\resizebox{0.5\textwidth}{!}{
\begin{tabular}{ c c c c}
\hline \hline \\
  & \multicolumn{2}{c}{Polarizability (in a.u.)}   \\
 \cline{2-3}  \\ Scheme &$^{1}$S$_{0}$     & $^{3}$P$_{0}$ & Excitation energy \\
 &&&(in cm\textsuperscript{-1 })  \\
 \hline \\
 Feller and Helgaker & 39.71(0.18) & 65.04(2.85) & 31577(33) \\\\
 Feller and Lesiuk  & 39.82(0.29) & 65.01(2.82) &31590(46)\\\\
 Martin   & 39.73(0.2) & 64.82 (2.63)&31578(34)\\\\
Peterson and Dunning  & 39.60(0.07) & 63.70(1.51) &31558(14)\\\\
Spread  & 0.22  & 1.34  & 32 \\
\hline \hline 
\end{tabular}
}
\label{table:5}
\end{table}

The next contributor to the uncertainty in the excitation energy and ground state polarizability is the incompleteness of one electron basis set. It is actually the biggest contributor to the uncertainty for the excited state polarizability. As one can see, the results have not converged even at the pentuple zeta basis set. So, one needs to extrapolate it to a CBS limit. There is no unique way to extrapolate it to the CBS limit. In this work, we have tested the four most popular approaches for basis set extrapolation. The first approach uses the three-parameter mixed exponential form suggested by Feller \cite{10.1063/1.462652} to extrapolate the DHF energy. The correlation energy is extrapolated using the Helgaker extrapolation scheme \cite{10.1063/1.473863}, which uses L\textsuperscript{-3 } error formula, with L as the highest angular momentum in the basis set. The second scheme uses the Feller formula for the DHF energy, and the correlation energy is extrapolated using the Lesiuk formula \cite{doi:10.1021/acs.jctc.9b00705}, which employs the Riemann zeta function to recover the missing energy contribution due to a truncated basis set. We have also tested the basis set extrapolation scheme of Martin \cite{MARTIN1996669}, which uses a Schwartz-type extrapolation formula to extrapolate both the DHF and correlation energies. The fourth approach advocated by Peterson and Dunning \cite{10.1063/1.466884} uses a cardinal dependent mixed exponential formula for both the DHF and correlation energies. 

Table \ref{table:5} presents the excitation energy and polarizabilities in various basis set extrapolation schemes. The Peterson and Dunning scheme gives the smallest excitation energy of 31558 cm\textsuperscript{-1 }. The Feller and Lesuik formula gives the highest excitation energy of 31590 cm\textsuperscript{-1 }. The ground state polarizability shows a small spread of 0.22 a.u.. However, the excited state polarizability gives a large spread of 1.34 a.u.. The Peterson-Dunning scheme gives the lowest polarizability of 63.70 a.u., whereas the Feller-Helgaker scheme gives us the highest excited state polarizability of 65.04 a.u.. We have chosen the results obtained from the Martin scheme for subsequent analysis here, as its predictions lie between the two extreme values for the polarizabilities and excitation energy.

\begin{table}[t]
\centering
\caption{Comparison of our computed results with the previous theoretical and experimental data.}
\resizebox{0.5\textwidth}{!}{
\begin{tabular}{ c c c c}
\hline \hline \\
  & \multicolumn{2}{c}{Polarizability (in a.u.)}   \\
 \cline{2-3}  \\ Method &$^{1}$S$_{0}$     & $^{3}$P$_{0}$ & Excitation energy \\
 &&&(in cm\textsuperscript{-1 })  \\
 \hline \\
 Dzuba and Derevianko \cite{Dzuba_2019} & 38.58 & 66.53 & 32348 \\
 Gropen and co-worker \cite{KEllingsen_2001}  & 39.13 & 66.50 &32707\\
 Ye and Wang \cite{PhysRevA.78.014502} &38.12 &67.69 &--\\
Singh and Sahoo \cite{PhysRevA.90.022511} &38.666(0.96) & --&--\\
Angom and co-worker \cite{PhysRevA.91.052504}  &38.75&--&--\\
Sahoo and co-worker \cite{PhysRevA.105.062815} &38.99(0.31)&-- &--\\
 
     Final Value    &  38.38 $\pm$ 0.37    & 64.81 $\pm$ 1.26    & 32343 $\pm$ 92 \\
 \\
 \\
 Experiment  & 38.8(8)\cite{PhysRevA.54.1973}  & --   & 32311\cite{DagGullberg_2000} \\
\hline \hline 
\end{tabular}
}
\label{table:6}
\end{table}

Visscher and co-workers \cite{PhysRevA.83.030503} have used an alternate strategy to estimate the basis incompleteness. They assumed that any property P (excitation energy or polarizability) calculated using the X$\zeta$,(X+1)$\zeta$ and (X+2)$\zeta$ basis sets would satisfy the relation $\frac{P_{(X+1)\zeta }-P_{(X)\zeta }}{P_{(X+2)\zeta }-P_{(X+1)\zeta }}=2$ for X$>$3. As the geometric series $\frac{1}{2}+\frac{1}{4}+\frac{1}{8}+\cdots $ converges to one, the missing contribution due to the basis set incompleteness to the excitation energy will be of the same order as the difference between the aug-cc-pVQZ basis and aug-cc-pV5Z basis set results; i.e. 24 cm\textsuperscript{-1}. Use of the same logic will lead to a basis set correction of 0.11 a.u. and 2.6 a.u. to the ground state and excited state polarizabilities over the values obtained using the aug-cc-pV5Z basis, which is consistent with the result of the extrapolation scheme suggested by Martin. The four-basis set extrapolation scheme leads to a spread of 32 cm\textsuperscript{-1} for the excitation energy and 0.22 a.u. and 1.34 a.u. for the ground and excited state polarizabilities. As the results in Martin's extrapolation scheme lie between the two extremes of the range, half of this spread can be considered uncertainty due to the incompleteness of the one-electron basis set. The maximum magnitude of the uncertainty for the frozen core approximation and higher-order relativistic effects can be taken to half of their total contributions. This leads to an uncertainty of 0.28 \% to the excitation energy and 0.97\%  and 1.94\% to the ground and excited state polarizabilities, respectively. Table \ref{table:4} tabulates the uncertainty associated with different corrections.

\begin{table}[t]
\centering
\caption{The BBR shift value, calculated by taking the difference between the static dipole polarizabilities obtained using the EOM-CC method and its comparison with the previously obtained theoretical result.}
\resizebox{0.5\textwidth}{!}{
\begin{tabular}{ c c c c}
\hline \hline \\
Zn atom & $\delta{\nu}$ (Hz)  & $\nu_{0}$ (Hz) & $\delta{\nu}/\nu_{0}$ \\
 \hline \\
This work  & (-0.227 $\pm$ 0.011)   & 9.68 $\times$ 10\textsuperscript{14 } & (-2.34 $\pm$ 0.11) $\times$ 10\textsuperscript{-16 } \\
\\
Previous work\cite{Dzuba_2019} & -0.244(10) & 9.68 $\times$ 10\textsuperscript{14 } & -2.51 $\times$ 10\textsuperscript{-16 } \\
(Scaled method) & & & \\
\hline \hline 
\end{tabular}
}
\label{table:7}
\end{table}

Our calculated results show very good agreement with the experimental values. The results are within the experimental error bar for ground state polarizability and show less than 0.1\% deviation from the experimental excitation energy value. There are no experimental results available for the excited state polarizability. Our recommended value for the excited state polarizability is 64.81 $\pm$ 1.26 a.u.. This value is slightly smaller than that recommended by Dzuba and Derevianko \cite{Dzuba_2019}, Ye and Wang \cite{PhysRevA.78.014502}, as well as Gropen and Coworker \cite{KEllingsen_2001} (see Table \ref{table:6}). None of these studies have reported uncertainties in their results, which makes it difficult to estimate the quality of the earlier predicted results. However, the study of Gropen and co-workers shows less favorable agreement with the experiment for both the excitation energy and ground state polarizability. Therefore, their prediction for the excited state polarizability is less trustworthy. The study by Dzuba and Derevianko shows results comparable to ours for the ground-state polarizability and excitation energy. Our method has the additional advantage that they are solely derived from the {\it ab initio} approach within a relativistic framework. These results are self-reliant, requiring no prior information of experimental data to predict the required properties. We believe that the values we obtained through the advanced relativistic EOM-CC method are highly reliable and can be confidently applied to estimate uncertainties in the high-precision measurements using the Zn atom. Furthermore, this method has the potential to be applied to other atomic clock candidates, enabling high-precision calculations across various atomic systems in the periodic table.

We would now like to employ the polarizability values obtained for the states participating in the clock transition using Eq. (\ref{eq:1}) to estimate the energy shift due to the BBR effect. The contribution from the dynamic correction $\eta$ is neglected as they are generally less than $7\times10\textsuperscript{-4}$ for the Zn atom \cite{Dzuba_2019}. Table \ref{table:7} presents the BBR shift value alongside previously reported data. The table clearly indicates that our estimated BBR shift to the clock transition $^{1}$S$_{0}$$\rightarrow$$^{3}$P$_{0}$ is smaller than the previously calculated value. Moreover, the relative BBR shift is very precise, and the uncertainty would affect the accuracy of the clock output in the 17$^{th}$ significant digit. The relative BBR shift for the Zn atomic clock will be half of the shift observed in the popular Sr atomic clock \cite{PhysRevResearch.3.L042036}.

\section{Conclusion}

We have calculated very precise electric dipole polarizability values for the ground and first excited states of the Zn atom using the relativistic equation-of-motion coupled-cluster theory. These values are immensely useful in estimating the BBR shift of the $^{1}$S$_{0}$$\rightarrow$$^{3}$P$_{0}$ clock transition in the above atom. The roles of finite basis size, electron correlation effects, and higher-order relativistic corrections to both the excitation energy and electric dipole polarizability values are analyzed systematically. Our study shows that incompleteness of the one-electron basis is a critical contributor to the error in the calculation, and the accuracy of the complete basis set result depends upon the choice of the basis set in the calculations. Our ground state polarizability value is well within the experimental error bar, and the excited state energy value is within 0.1\% of the experimental value. This demonstrates the potential of our employed method to estimate the properties of atomic systems, the zinc atom in particular, precisely. Based on these analyses, we provide the recommended value for the excited state dipole polarizability to be 64.81 $\pm$ 1.26 a.u., which is lower than the earlier values reported without any error bars. This value leads to a smaller BBR shift compared to the earlier estimation. 
Our study carries significant importance for the optical clock experiment pertaining to the zinc atom, and our result shows that the Zn atom can be an appealing alternative to optical atomic clocks already in use. The current protocol is versatile enough to be extended to other atomic clock candidates to estimate their BBR shifts with similar accuracy.

\section*{Acknowledgment}

The authors acknowledge the support from the ISRO for financial support under its RESPOND program, CRG and Matrix project of DST-SERB, CSIR-India, DST-Inspire Faculty Fellowship, Prime Minister's Research Fellowship, IIT Bombay super computational facility, and C-DAC Supercomputing resources (PARAM Yuva-II, Param Bramha) for computational time.

\section*{Conflict of interest}

The authors declare no competing financial interest.

%

\end{document}